\documentclass[floatfix,twocolumn,amsmath,amssymb,aps,prb,showpacs,letter,superscriptaddress]{revtex4}
\usepackage{latexsym,epsfig,bm,times,psfrag,subfigure}
 \usepackage{graphicx}
\usepackage{color}

\newcommand{\hto}{Ho$_{2}$Ti$_{2}$O$_{7}$}

\newcommand{\dto}{Dy$_{2}$Ti$_{2}$O$_{7}$}
\newcommand{\eto}{Er$_{2}$Ti$_{2}$O$_{7}$}

\begin{document}
\title{Order-by-disorder in the XY pyrochlore antiferromagnet revisited}
\author{Pawel Stasiak}
\affiliation{Department of Physics and Astronomy, University of Waterloo, Waterloo, ON, N2L 3G1, Canada.}
\affiliation{Department of Mathematics, University of Reading, Whiteknights, PO Box 220, Reading RG6 6AX, UK.}
\author{Paul A. McClarty}
\affiliation{Department of Physics and Astronomy, University of Waterloo, Waterloo, ON, N2L 3G1, Canada.}
\affiliation{Max Planck Institute for the Physics of Complex Systems, N\"{o}thnitzer Str. 13, Dresden, 01187, Germany.}
\author{Michel J. P. Gingras}
\affiliation{Department of Physics and Astronomy, University of Waterloo, Waterloo, ON, N2L 3G1, Canada.}
\affiliation{Canadian Institute for Advanced Research, 180 Dundas Street West, Suite 1400, Toronto, ON, M5G 1Z8, Canada.}
\date{\today} 
\begin{abstract}
We investigate the properties of the XY pyrochlore antiferromagnet with local
$\langle 111 \rangle$ planar anisotropy. We find the ground states and show
that the configurational ground state entropy is subextensive. By computing the free energy
due to harmonic fluctuations and by carrying out Monte Carlo simulations, we
confirm earlier work indicating that the model exhibits thermal order-by-disorder 
leading to low temperature long-range order consisting of discrete
magnetic domains. We compute the spin wave spectrum and show that thermal and
quantum fluctuations select the same magnetic structure. 
Using Monte Carlo simulations, we find that the state selected by
thermal fluctuations in this XY pyrochlore antiferromagnet
can survive the addition of sufficiently weak 
nearest-neighbor pseudo-dipolar
 interactions to the spin Hamiltonian.
We discuss our results in relation to the \eto\ pyrochlore antiferromagnet.

\end{abstract}


\maketitle

\section{Introduction}

The geometric frustration of magnetic interactions on lattices of magnetic
moments often leads to a configurational classical ground state entropy that scales with the volume
of the system, $V$, as $V^{\alpha}$ with $0<\alpha\leq 1$. This can have some
unusual consequences. A well known example is the 
Ising model on the triangular lattice with nearest-neighbor antiferromagnetic interactions 
which has an extensive ground state entropy and exhibits no finite temperature
transition. \cite{triangular_Ising_AF} While in real materials
a $V^{\alpha}$ entropy left by the leading interactions is often 
energetically lifted by weaker interactions, 
leading to long-range magnetic order, there are some exceptions. For example,
in the  \dto\ and \hto spin ice materials,~\cite{SpinIce}  in which the magnetic moments 
are described by Ising spins,
the extensive ($\alpha=1$) low temperature entropy caused by frustration of the leading
effective ferromagnetic nearest-neighbor interactions is indeed lifted by
the perturbing long-ranged part of the dipolar interaction.~\cite{Gingras-CJP,DSI_LRO,Isakov-rules} 
However, the degeneracy lifting in this system is so weak that the
theoretically expected phase transition to
magnetic long-range order is inhibited by a freezing into a 
spin ice state without long-range order. \cite{denHertog, freezing}

Another possibility in a system with an exponentially ($\exp[C V^{\alpha}]$)
large number of classical
degenerate ground states is that thermal or quantum fluctuations might
select a subset of states about which the density of zero modes is
greatest. These entropic and quantum state selection mechanisms are both
referred to as order-by-disorder.~\cite{OBDmodel1,OBDmodel2,OBDmodel3,Footnote} 
Among pyrochlore antiferromagnets, 
in which the spins sit on a lattice of corner-sharing tetrahedra,
Moessner and Chalker have given a
criterion for the 
occurrence of long-range order induced by thermal fluctuations.~\cite{MoessnerChalker} 
 This criterion is based on the degree of divergence of the statistical weight of particular spin
configurations $-$ a power-counting argument depending on the number of zero energy excitations 
(zero modes) for a given spin
configuration and the number of dimensions of the ground state manifold. For
example, this criterion indicates that the XY antiferromagnet with globally
coplanar spins (spins perpendicular to the global $[001]$ axis) should
exhibit entropic selection $-$ a result which is borne out by Monte Carlo
simulations. \cite{MoessnerChalker} For such XY systems, this comes about because the number of
zero modes about collinear spin configurations is proportional to the number
of spins whereas the configurational 
entropy in the ground state is subextensive, growing as $V^{2/3}$.~\cite{MoessnerChalker,OBD1} 

This article is concerned with the pyrochlore XY antiferromagnet
with local $\langle 111\rangle$ spins meaning that there is a different easy
plane for each of the four tetrahedral sublattices.~\cite{OBD1,ETO,OBD2,OBD3} 
Because such a model
preserves the cubic symmetry of the pyrochlore lattice and because the single
ion crystal field can, and does in various materials,~\cite{Review} generate such an anisotropy, 
it is more physical than the aforementioned 
pyrochlore XY model with a global easy axis.~\cite{MoessnerChalker} 
The model has been recognized to exhibit a continuous
degeneracy in its classical ground state.~\cite{OBD2,OBD3} 
Monte Carlo simulations of the local 
$\langle 111\rangle$ XY antiferromagnet \cite{OBD1,OBD2,OBD3} indicate that
it exhibits two phases $-$ a high temperature paramagnetic phase and a low
temperature long-range ordered phase. We refer to the magnetic structure in
the ordered phase as $\psi_{2}$ to be consistent with Ref.~[\onlinecite{ETO3}] 
and the group theory literature. A calculation of the spectrum of the
Hessian about different discrete ground states \cite{OBD2} suggests that the
observed long-range ordered spin configuration in Monte Carlo simulations
has the largest density of zero
modes of all the degenerate ground states and, consequently, that the observed
transition is an example of classical (thermal) OBD.~\cite{OBD2,OBD3}
However, it has been suggested that selection of the long-range ordered $\psi_2$ state 
might not survive in the thermodynamic limit.~\cite{OBD2}

In this article, we give a systematic account of the properties of the local 
$\langle 111\rangle$ XY pyrochlore. We present in Section~\ref{sec:model} the model
and its ground states. 
In Section~\ref{sec:methods}, we discuss some of the details of the Monte Carlo
simulations performed in this work.
Section~\ref{sec:OBD} reports results of an analytical
and numerical investigation of the thermal order-by-disorder
mechanism, providing strong evidence that the fluctuation selection 
mechanism of the $\psi_2$ state
does survive in the thermodynamic limit and giving  further insight
into its physical origin.  We also include 
in Section~\ref{sec:OBD} a subsection showing that
there is a quantum order-by-disorder mechanism in the XY model as speculated
but not shown in Ref.~[\onlinecite{ETO}]. Finally, in Section~\ref{sec:eto}, we
describe the material \eto\ which is an easy plane antiferromagnet exhibiting
the $\psi_2$ structure in its ordered phase \cite{ETO,ETO3} 
and which provides an experimental
motivation for studying this model. In particular, we discuss the effect of weak
dipolar interaction on the XY antiferromagnet and the problem this interaction
poses for understanding the long-range ordered phase of \eto\ with 
$\psi_2$ structure.~\cite{ETO,ETO3}

\section{Model}
\label{sec:model}

In this work, we mostly focus on 
the problem of the zero and finite temperature behavior of interacting classical spins
of length $\vert {\mathbf{S}}\vert = 1$ on the sites of a pyrochlore lattice of corner-sharing
tetrahedra with an infinite single-ion anisotropy such that the spins lie within
their respective local XY planes perpendicular to the local 
$\langle 111\rangle$ directions. 
In Section \ref{sec:OBDquantum}, we discuss the problem of order-by-disorder
due to quantum fluctuations in a model with spin operators ${\mathbf{S}}$.~\cite{OBD1,ETO,OBD2}

The interactions are taken to be antiferromagnetic
isotropic exchange between nearest neighbors with coupling $J$ ($J>0$).
 Later on we also consider, as a perturbation, pseudo-dipolar 
interactions solely between nearest neighbours 
and with coupling strength $\mathcal{D}$.
Thus the Hamiltonian is taken to be
\begin{equation} H = J\sum_{\langle i,j\rangle} \mathbf{S}_{i}\cdot\mathbf{S}_{j} + \mathcal{D}R_{\rm
  nn}^{3}\sum_{\langle i,j\rangle} \frac{\mathbf{S}_{i}\cdot\mathbf{S}_{j}}{|\mathbf{R}_{ij}|^{3}} 
- \frac{3(\mathbf{S}_{i}\cdot\mathbf{R}_{ij})(\mathbf{S}_{j}\cdot\mathbf{R}_{ij})}{|\mathbf{R}_{ij}|^{5}}
  \label{eqn:model}  \end{equation}
where $R_{\rm nn}^{3}$ is the nearest-neighbor distance.

Consider first the exchange-only model with $\mathcal{D}=0$.
In this case, the Hamiltonian in Eq. (\ref{eqn:model})
can be put into the form $H_{\rm ex} 
= J\sum_{\rm t} \left( \mathbf{S}_{\rm t}^{2}  - 4S^{2} \right)$,
 where the sum runs over all connected tetrahedra \cite{MoessnerChalker} and $\mathbf{S}_{\rm t}$ is
the total spin on each tetrahedron. It follows that the ground states are all
those states with zero net magnetic moment 
($\mathbf{S}_{\rm t}=0$) on each tetrahedron. Therefore, we write down the
conditions for the three components of the total moment on a tetrahedron to be
zero. In doing so, we impose the XY constraint so that the orientation of spin
$a$, (for sublattices $a=1,2,3,4$), is given by a single angle $\phi_{a}$ measured with respect
to axes within the local plane (normal to the relevant local $[111]$ direction)
 given in Ref.~[\onlinecite{ETO6}].
 The condition of zero moment on each tetrahedron  can then be written as
\begin{align*}
&\cos\left(\phi_{1}\right) + \cos\left(\phi_{2}\right) 
= \cos\left(\phi_{3}\right) + \cos\left(\phi_{4}\right) \\
&\cos\left(\phi_{1}'\right) + \cos\left(\phi_{3}'\right) 
= \cos\left(\phi_{2}'\right) + \cos\left(\phi_{4}'\right) \\
&\cos\left(\phi_{1}''\right) + \cos\left(\phi_{4}''\right) 
= \cos\left(\phi_{2}''\right) + \cos\left(\phi_{3}''\right),
\end{align*}
where $ \phi_{a}'\equiv  \phi_{a}+\frac{2\pi}{3}$ 
and $ \phi_{a}''\equiv  \phi_{a}+\frac{4\pi}{3}$.
There are four solution branches to these equations.
Each branch corresponds to a continuous degeneracy wherein all four spins are rotated
smoothly within their respective local $[111]$ XY plane. 

We place an overbar on $\phi_{a}$ ($\bar \phi_a$) to signify the angle for 
sublattice $a$ giving an energy minimum (zero moment on each tetrahedron).
Then, we label these branches in the following way:
\begin{align}
{\rm Branch \hspace{2pt}1:} \hspace{6pt} \bar\phi & \equiv \bar \phi_{1} = \bar \phi_{2} = \bar \phi_{3} = \bar \phi_{4} \notag \\
{\rm Branch \hspace{2pt}2:} \hspace{6pt} \bar\phi & \equiv \bar \phi_{1} = \bar \phi_{2} = -\bar \phi_{3} = -\bar \phi_{4} \notag \\
{\rm Branch \hspace{2pt}3:} \hspace{6pt} \bar\phi & \equiv \bar \phi_{1} = \bar \phi_{3}\hspace{1pt}, 
\hspace{4pt} \frac{2\pi}{3} - \bar\phi = \bar \phi_{2} = \bar \phi_{4} \notag \\
{\rm Branch \hspace{2pt}4:} \hspace{6pt} \bar \phi & \equiv \bar \phi_{1} = \bar \phi_{4}\hspace{1pt}, \hspace{4pt} \frac{4\pi}{3} - \bar \phi = \bar \phi_{2} = \bar \phi_{3}. \label{eqn:groundstates}
\end{align}
A further discussion of these solutions can be found in Appendix A.

To enumerate all the ground states on the pyrochlore lattice we
first tile all the tetrahedra with a particular spin configuration from Branch
$1$. Then, we choose a line of nearest-neighbor spins traversing the length
$L$ of the system. The sublattice labels of the spins on the chain alternate
between two values $a$ and $b$. There are six such pairs of labels. One can then
transform the spins along the chain so that the spin configurations of the
associated tetrahedra belong to another branch of solutions. 
For example, consider a single chain made of
sublattices $3$ and $4$. All the local angles along
this chain are identical initially and equal to, say $\theta$. We can
transform these to $-\theta$ with no energy cost. 
Therefore, the entropy within the ground state manifold scales as $L^{2}$ as first noted in
Ref.~[\onlinecite{OBD1}]. This is in contrast to both the Heisenberg 
pyrochlore antiferromagnet and the global easy axis (Ising) pyrochlore
antiferromagnet both of which have an extensive entropy.

We note that the four branches in Eq.~(\ref{eqn:groundstates}) intersect in
pairs. These intersection points are at 
special sublattice angles $\bar\phi=n\pi/3$ with integer $n$. 
We refer to these as $\psi_{2}$ states in the rest of this
article. By exploiting these intersection angles to move between the branches, one can smoothly visit all
the ground states on a single tetrahedron and, indeed, on the whole pyrochlore
lattice. If we return to the above chain of sublattices $\#3$ and $\#4$,
the $\bar\phi=0$ configuration allows the tetrahedra along this 
chain to pass smoothly from Branch $1$ to Branch $2$.
As shown in Refs.~[\onlinecite{ETO},\onlinecite{OBD2}] and in Section~\ref{sec:OBD} below, 
thermal fluctuations have
the effect of selecting a magnetic structure with  $\mathbf{q}=0$ ordering wavevector and spin
orientations at these discrete $\bar\phi$ angles. 
There are six distinct $\psi_{2}$ ground states which are the
six $\mathbf{q}=0$ ordered states with tetrahedra tiled with local angles 
$\bar \phi_a=n\pi/3$ for sublattices $a=1,2,3,4$ and with integer $n$.
One can take the observation that the lattice zero modes are along sublattice 
chains to understand an aspect of the Monte Carlo results of Ref.~[\onlinecite{OBD2}]; 
in particular, the finite-size
scaling of the average energy of the $\psi_2$ states at low temperature.
 Since this point is somewhat removed from the main story of the paper, 
we present the argument in Appendix~\ref{sec:AvE}.


When $\mathcal{D}\neq 0$, the continuous ground state degeneracy of the exchange only
model is ``immediately''
replaced with a discrete global degeneracy with $\mathbf{q}=0$ ordering
wavevector selected from the manifold of states described above. 
These energetically selected states are
 referred to as  the $\psi_{4}$ states ~\cite{ETO3} or
 Palmer-Chalker states in the literature after Ref.~[\onlinecite{PalmerChalker}].
 The angles specifying the $\psi_{4}$ states are 
\begin{align}
{\rm State \hspace{2pt}1:}\hspace{6pt} & \bar \phi_{1} = \bar  \phi_{2} = \frac{\pi}{2} & 
\hspace{0.5cm} \bar \phi_{3} = \bar \phi_{4} =
\frac{3\pi}{2} \notag \\
{\rm State \hspace{2pt}2:}\hspace{6pt} &  \bar \phi_{1} = \bar \phi_{4} = \frac{7\pi}{6} & 
\hspace{0.5cm}  \bar \phi_{2} = \bar \phi_{3} =
\frac{\pi}{6} \notag \\
{\rm State \hspace{2pt}3:}\hspace{6pt} &  \bar \phi_{1} = \bar \phi_{3} = \frac{11\pi}{6} & 
\hspace{0.5cm} \bar \phi_{2} = \bar \phi_{4} =
\frac{5\pi}{6} \label{eqn:psi4states}
\end{align}
and the time-reversed configurations. 
In anticipation of what follows in  Section~\ref{sec:eto},
we note that the $\psi_4$ states 
are the ground states one finds for antiferromagnetic nearest-neighbor 
exchange with sufficiently weak nearest-neighbor pseudo-dipolar as well as for true $1/r^3$ 
long-range magnetostatic dipolar interactions in the classical Heisenberg 
pyrochlore antiferromagnet model.~\cite{PalmerChalker,SpinWaves,Maestro_PRB}
Interestingly, the $\psi_4$ states are found experimentally to be the ground
state of the Gd$_2$Sn$_2$O$_7$  pyrochlore antiferromagnet,~\cite{yes-GSO}
 but not of the closely related Gd$_2$Ti$_2$O$_7$ material.~\cite{not-GTO}

\section{Monte Carlo Method}
\label{sec:methods}

In Sections~\ref{sec:OBDb} and \ref{sec:eto} below, we report results from
Monte Carlo simulations of the local $\langle 111\rangle$ XY
pyrochlore antiferromagnet. In this section, we give details of the Monte
Carlo algorithm and the observables that were measured in the Monte Carlo simulations.

The Monte Carlo simulations were performed using parallel tempering \cite{PT} in which
$N_{T}$ replicas of a system of $N$ spins, each at a different temperature and with a different series
of pseudo-random numbers, are simulated simultaneously. In addition to local
spin moves, parallel tempering swaps that exchange configurations
between a pair of temperatures are attempted. The configuration swap attempts are
accepted or rejected based on a Metropolis condition that preserves detailed
balance. Parallel tempering has been shown, in systems known to equilibrate
slowly using other methods, to improve performance substantially. \cite{PT}
Replica swaps are attempted with a frequency of one attempt every $100$ local
Monte Carlo sweeps. A local Monte Carlo sweep consists of $N$ spin move attempts.
 In our simulations, $N_{T}=64$ with
either a constant increment between the temperatures, or with the temperatures
self consistently adjusted to obtain a uniform parallel tempering acceptance
rate. 

With each spin carrying a single angular coordinate $\phi_{i}$, the local spin
moves involve choosing an angle increment $\delta\phi_{i}$ from a uniform
distribution between $-\delta\phi_{\rm max}$ and $\delta\phi_{\rm max}$. 
The
angle of spin ${\mathbf{S}}_i$  was updated to $\phi_{i}+\delta\phi_{i}$ and each
tentative spin rotation was accepted or rejected based on a Metropolis test. 
The
maximum increment $\delta\phi_{\rm max}$ was updated every $100$ Monte Carlo
moves in order to maintain the spin move acceptance rate at $50\%$.

Physical observables were computed every $100$ Monte Carlo sweeps. 
To determine the presence of long-range order with ordering
wavevector $\mathbf{q}=0$ (expected for sufficiently small ${\cal D}/J$ in the model discussed
above \cite{PalmerChalker,SpinWaves,Maestro_PRB}), 
the sublattice magnetization was computed \cite{OBD1}

\begin{equation} M_{4} = \left\langle \sqrt{\frac{1}{4}\sum_{a=1}^{4}\left(
    \frac{1}{N_{P}}\sum_{i=1}^{N_{P}} \mathbf{S}_{i,a} \right)^{2}} \right\rangle_{\rm th},
\end{equation}
where each spin carries an fcc lattice label $i$ and a sublattice label
$a$ (see Ref.~[\onlinecite{Enjalran}]) and the number of sites in the lattice is $N\equiv
4N_{P}$, where $N_{P}$ is the number of fcc sites.  The angled brackets
$\langle\ldots\rangle_{\rm th}$ denote a thermal average. 
In order to distinguish the $\psi_{2}$ (Refs.~[\onlinecite{ETO,ETO3}])
 				and $\psi_{4}$ (Refs.~[\onlinecite{ETO3,PalmerChalker}]) phases, 
we introduce unit vectors $\mathbf{\hat{e}}_{a}^{(\gamma(d))}$ which are oriented in the expected spin
directions on each sublattice $a$ for magnetic structure 
identified by the label $\gamma$ 
 with the domains labelled $d$, for both the $\gamma=\psi_2$ and $\gamma=\psi_4$ structures. 
From the combination
\begin{equation}
 \Psi^{(\gamma(d))} =
 \frac{1}{N_{P}}\sum_{i=1}^{N_{P}}\sum_{a=1}^{4} \mathbf{S}_{i,a}\cdot\mathbf{\hat{e}}_{a}^{(\gamma(d))}, 
\label{Psi_def}
\end{equation}
we compute the order parameter
\begin{equation} q_{\gamma} = \left\langle \sum_{d} \left( \Psi^{(\gamma(d))}
  \right)^{2} \right\rangle_{\rm th}, 
\label{qgamma}  
\end{equation}
for the $\gamma=\psi_{2}$ (Refs.~[\onlinecite{ETO,ETO3}]) 
and $\gamma=\psi_{4}$ (Refs.~[\onlinecite{ETO3,PalmerChalker}])
magnetic structures.
In Eq.~(\ref{qgamma}), the  sum is taken over 
a choice of three out of the six magnetic domains, for each of these two structures,
which are not related to one another by time reversal. 
The spin directions corresponding to the domains for $\psi_4$ and $\psi_2$ 
are given in Section II. $-$ for $\psi_4$ in Eq.~(\ref{eqn:psi4states}) 
and for $\psi_2$ we have all angles $\bar{\phi}_a=n\pi/3$. 
The order parameter for $\psi_{4}$, $q_{\psi_{4}}$, is the same one computed in
the simulations of Ref.~[\onlinecite{Heisenberg}]. The magnetic specific heat per spin 
was computed from the fluctuations in the total energy of the system.

The sensitivity of the results to the initial spin configurations was assessed
by comparing the results of simulations starting from (i) random
configurations with a different configuration for each thermal replica, (ii)
$\psi_{2}$ ordered states and (iii) $\psi_{4}$ ordered states. To ensure that
equilibration was reached for each simulation, we checked that the results were independent
of initial conditions. Also, the evolution of the order parameters
was monitored during the course of each simulation to ensure that they
reached a stationary state before the statistics were collected. 
Equilibration issues are discussed further in Sections~\ref{sec:OBDb} and \ref{sec:eto}.

\section{Order-by-disorder}
\label{sec:OBD}

In this section, we consider the exchange-only model ($\mathcal{D}=0$) given in Eq.~(\ref{eqn:model}). 
General arguments given in Ref.~[\onlinecite{MoessnerChalker}] indicate 
that the XY antiferromagnet with coplanar spins should exhibit a thermally 
driven order-by-disorder transition. This argument does not straightforwardly carry over to the
noncoplanar $\langle 111\rangle$ XY antiferromagnet. 
Simulation evidence for thermal order-by-disorder in the local $\langle 111\rangle$ XY model transition was presented
in Refs.~[\onlinecite{OBD1,ETO,OBD2,OBD3}]. The classical degeneracies of this model
were identified in Ref.~[\onlinecite{OBD2}] and order-by-disorder was found via
Monte Carlo simulations. 
However, the possibility was mentioned in Ref.~[\onlinecite{OBD2}] that the 
temperature at which long-range order with a nonzero $\psi_2$ order parameter develops
might vanish in the thermodynamic limit. 
We present simulation results which provide compelling evidence that, 
for the exchange-only (${\cal D}=0$) model,  
a first order phase transition to a long-range ordered $\psi_2$ state 
persists in the thermodynamic limit.
We begin, however, with a previously unreported calculation of the
free energy including only harmonic fluctuations which exposes a thermal
 $\psi_2$ order-by-disorder in the thermodynamic limit. 
Then, having investigated the order-by-disorder mechanism in the exchange-only (${\cal D}=0$) model, 
we discuss in Section~\ref{sec:eto} the effect of competing
nearest-neighbor pseudo-dipolar (${\cal D}\ne 0$) interactions in this model.

\subsection{Computation of the free energy}
\label{sec:OBDa}

In this section, we show that certain discrete spin configurations from the
manifold of $\mathbf{q}=0$ ground states minimize the free energy computed
from harmonic fluctuations about the classical ground states. We assume that
every tetrahedron on the lattice is tiled with the same spin configuration
({\it i.e.} that the ordering wavevector is $\mathbf{q}=0$).
 If we constrain the ordering to be $\mathbf{q}=0$, the spin configuration 
is fixed by specifying four angles $\phi_a$ - one for each sublattice. 
Let the angles in a ground state configuration be
denoted $\bar{\phi}_{a}$ for which the ground state energy is $H({\bar{\phi}_{a}})=NE_{g}$, where
$N$ is the number of spins. 
We then we consider small fluctuations $\delta\phi_i$ about these
angles $\phi_{i}=\bar{\phi}_{i}+\delta\phi_{i}$.
The terms linear in $\delta\phi_{i}$ vanish, so the Hamiltonian $H = NE_{g} + H_{2} + \ldots$,
where $H_{2}$  is the part harmonic in the angular deformations. 
$H_2$ is written in $\mathbf{k}$ space as
$H_{2}=\sum_{\mathbf{k},a,b} 
\delta {\phi}_{a} (\mathbf{k}) 
A^{ab}(\mathbf{k}) 
\delta {\phi}_{b}(-\mathbf{k})$ .
Here, $\delta {\phi}_{a}(\mathbf{k})=(1/\sqrt{N_{P}})\sum_{\mathbf{R}_{\mu}}
\exp(i\mathbf{k}\cdot(\mathbf{R}_{\mu}+\mathbf{r}_{a}))\delta\phi_{a}(\mathbf{R}_{\mu})$,
where $\mathbf{R}_{\mu}$ are the fcc lattice points and $\mathbf{r}_{a}$ are
the vectors for the tetrahedral basis (see Ref.~[\onlinecite{Enjalran}] for notation convention). 
This choice of convention for the lattice labelling ensures that the Hessian
$A^{ab}(\mathbf{k})$ is real. 
The eigenvalues $\lambda_{A}(\mathbf{k})$ of $A^{ab}(\mathbf{k})$ are
nonnegative, reflecting the stability of the ground states. The spectrum of
$A^{ab}$ is computed in $\mathbf{k}$ space as a function of the ground
state for each branch.

One finds that for
the special minimum energy configurations
$\bar{\phi}_{a}=n\pi/3$ for $a=1,2,3,4$,
the four eigenvalues, $\lambda_{A}(\mathbf{k})$, ($A=1,2,3,4$) of the 
$A^{ab}(\mathbf{k})$ Hessian take the form
\begin{align*} \bar{\phi}_{a}&=0,\pi & \lambda_{A} = 1\pm \cos(\mathbf{k}\cdot\mathbf{r}_{12}),
  \hspace{5pt} & 1\pm
  \cos(\mathbf{k}\cdot\mathbf{r}_{34}) \\ 
  \bar{\phi}_{a}&=\pi/3, 4\pi/3 & 
\lambda_{A} = 1\pm \cos(\mathbf{k}\cdot\mathbf{r}_{13}), \hspace{5pt} & 1\pm
  \cos(\mathbf{k}\cdot\mathbf{r}_{24}) \\
  \bar{\phi}_{a}&=2\pi/3, 5\pi/3 & 
\lambda_{A} = 1\pm \cos(\mathbf{k}\cdot\mathbf{r}_{23}), \hspace{5pt} & 1\pm
  \cos(\mathbf{k}\cdot\mathbf{r}_{14}) 
\end{align*}
where each row gives the 
 four eigenvalues for the indicated particular set of  $\bar \phi_a$ 
($a=1,2,3,4$) angles and which correspond to the aforementioned $\psi_{2}$ states.
 The vector $\mathbf{r}_{ab}$ joins nearest neighbors with
sublattice labels $a$ and $b$.  The $\psi_{2}$ states are distinguished from
the other ground states in having a much higher density of zero modes $-$ $2$
planes of zero modes in the first Brillouin zone. \cite{OBD2,OBD3}
 The planes come about because, at these angles, one can smoothly introduce defects into
the system along (what are usually referred to as $\alpha$ and $\beta$)
 chains in the crystal as described in Section~\ref{sec:model}. 
At angles away from the $\psi_{2}$ states, one can
still introduce chain defects but not continuously.

\begin{figure}
\includegraphics[width=\columnwidth]{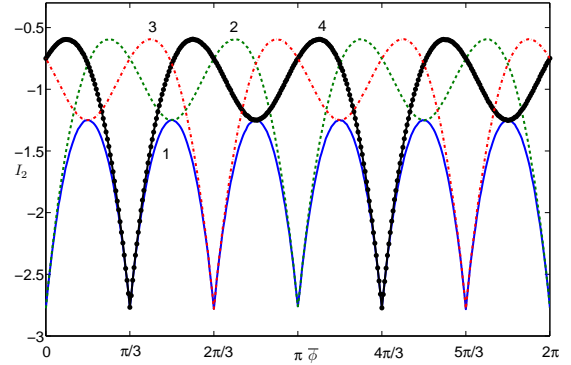}
\caption{(Color online) Plot showing the harmonic free energy contribution
  $I_{2}$ for each of the four branches 
($A=1,2,3,4$) of ground states with 
the $\bar \phi_a$ angle for each sublattice $a=2,3,4$  expressed in terms of $\bar \phi_1$ via
the parametrization given in Eq.~(\ref{eqn:groundstates}).
Each of the four curves is labelled by its ground state branch number taken from
  Eq.~(\ref{eqn:groundstates}). Note therefore that the horizontal axis does
  not identify a unique spin configuration but rather a unique configuration
  for each of the four branches. The minima in the free energy appear
  where pairs of ground state branches meet - at the $\psi_{2}$ spin configurations.}
\label{fig:FE}
\end{figure}

We now compute the free energy for each $\mathbf{q}=0$ configuration after
dropping all terms in the Hamiltonian beyond the harmonic terms. The free
energy $F[{\bar{\phi}}]$ at inverse temperature, $\beta$, is given at the harmonic
level by
\begin{equation} F[{\bar{\phi}}] \approx NE_{g} -\frac{1}{\beta}\log \left[ \left( \prod_{a,\mathbf{k}}\int
  d[\delta\phi_{a}(\mathbf{k})] \right) \exp\left( - \beta H_{2} \right)
  \right]   
\end{equation}
and hence
\begin{equation} F[{\bar{\phi}}] \approx NE_{g} - \frac{N}{2\beta} \log \left(\frac{\pi}{\beta}
  \right) + \frac{1}{2\beta} \sum_{\mathbf{k}} \log \left( {\rm det}
  A(\mathbf{k})  \right). 
\label{eqn:FE} 
\end{equation}
In the limit as $N\rightarrow \infty$, the third 
(last) term on the right hand side of
Eq.~(\ref{eqn:FE}) is $\frac{N}{4}\frac{1}{4}\frac{1}{(2\pi)^{3}} I_{2}$ 
with $I_{2}=\sum_{A}\int d^{3}\mathbf{k} \log \left(
  \lambda_{A}(\mathbf{k})  \right)$ for the four ground state branches given
  in Section~\ref{sec:model} 
(see Appendix \ref{app:factors_of_four} for
a discussion of the various prefactors of $I_2$). 
We evaluate the integral numerically using
a Monte Carlo method with
  $10^{8}$ points, noting that the singularities for the $\psi_{2}$
  spin configurations are integrable because the integrals take the form
  $\int_{0}^{c} dk \log k$ for some constant $c$.~\cite{integrable} 
The results are shown in Fig.~\ref{fig:FE}. Evidently, $I_2$, and consequently
the free energy $F[{\bar{\phi}}]$, is minimized for the $\psi_{2}$ states at
  $\bar{\phi}_{a}=n\pi/3$ where pairs of branches meet.

It is often the case that one can simulate the effect of order-by-disorder by
introducing a term into the Hamiltonian of the form
\begin{equation}
 H_{\rm OBD} = -|\Gamma|\sum_{i,j} 
\left(\mathbf{S}_{i}\cdot\mathbf{S}_{j}\right)^{2},
\label{eqn:collinear} 
\end{equation}
that selects the most collinear spin
configurations among the classical ground states 
(see, for example, Refs.~[\onlinecite{OBDmodel1,OBDmodel2,OBDmodel3}]).
The usual argument for the selection of such states is that collinear spin
configurations have, among all states, fluctuations that couple most
strongly because fluctuations are responsible for effective fields perpendicular to
the spin direction even in the broken symmetry phase. The local XY and zero moment constraints of the XY pyrochlore
antiferromagnet ensure that the spins cannot be collinear, but it is interesting to ask
whether the $\psi_{2}$ configurations are the most collinear states within the set of ground
states. One finds that Eq.~(\ref{eqn:collinear}) is constant within the whole ground state manifold of
Eq.~(\ref{eqn:groundstates}). However, the sum $-\sum_{i,j}
  \left|\mathbf{S}_{i}\cdot\mathbf{S}_{j}\right|$
is minimized by the $\psi_{2}$ states which lends some credence to the
intuition that the most collinear states among 
all the classically degenerate zero temperature ground states must be selected.

\subsection{Quantum selection}
\label{sec:OBDquantum}

\begin{figure}
\includegraphics[width=\columnwidth]{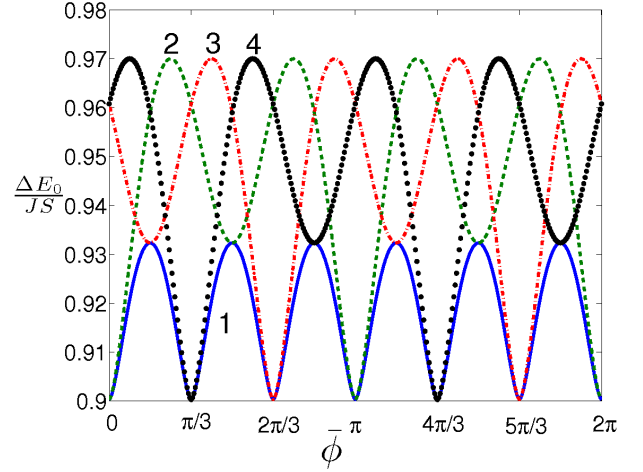}
\caption{(Color online) Zero point energy for the pyrochlore XY model from the linear spin
  wave spectrum computed from the classical ground states. The horizontal axis
  is the $\phi$ parameter given in Eq.~(\ref{eqn:groundstates}) and each curve
  carries a label identifying the branch of ground states to which it belongs.  The quantum
  correction to the classical ground state energy is minimized for the
  $\psi_{2}$ states.}
\label{fig:ZP}
\end{figure}

Having shown that thermal fluctuations  select the $\psi_{2}$ states, we
now turn to the effect of quantum fluctuations which, in general, need not select the same states. 
In this section, we present the spin wave spectrum computed using the
Holstein-Primakoff transformation treated in a large $S$ expansion and
truncated at harmonic order. The calculation is performed for the Hamiltonian
in Eq.~(\ref{eqn:model}) with $\mathcal{D}=0$. In the coordinate system with $z$
axes taken along the $\langle 111\rangle$ directions, the Hamiltonian is
written $H=\sum \mathcal{J}_{ij}^{\alpha\beta}S_{i}^{\alpha}S_{j}^{\beta}$
where $\alpha$ and $\beta$ denote the spin components. The local Ising components of the
matrix of interactions $\mathcal{J}^{\alpha\beta}_{ij}$ are set equal to zero $-$
 this imposes a soft XY constraint because the computation of the spin wave spectrum implicitly allows
fluctuations out of the easy planes. 
This is relevant to the \eto\ pyrochlore antiferromagnet
whose single-ion crystal field doublet,
characterized by an anisotropic $g$-tensor
with two eigenvalues such that $g_\perp > g_\parallel$,
allows for a description in terms of an effective spin-1/2 model.~\cite{Yb2Ti2O7-example}
Working in reciprocal space with $N$ spins and $N_{P}$ primitive lattice sites, one rewrites the spin Hamiltonian
in terms of boson operators, with \cite{SpinWaves,Maestro_PRB}
\begin{align} 
\tilde{S}_{a}^{z}(\mathbf{k})
& =\sqrt{N_{P}}S\delta_{\mathbf{k},0}e^{-i\mathbf{k}\cdot\mathbf{r}_{a}} 
-\frac{1}{\sqrt{N_{P}}}a_{a}^{\dagger}(\mathbf{k}')a_{a}(\mathbf{k}'-\mathbf{k}) \\
\tilde{S}_{a}^{x}(\mathbf{k}) & =\sqrt{\frac{S}{2}} \left( a_{a}^{\dagger}(\mathbf{k}) + a_{a}(-\mathbf{k}) \right) \\
\tilde{S}_{a}^{y}(\mathbf{k}) & = i \sqrt{\frac{S}{2}} \left( a_{a}^{\dagger}(\mathbf{k}) - a_{a}(-\mathbf{k}) \right),
\end{align}
on each site with the new  $z$ axis
now taken to be the quantization axis. The
quantization axis is taken within the ground state manifold of the model with
antiferromagnetic exchange as parametrized by local angles $\bar{\phi}_{a}$ for
$a=1,2,3,4$ given in Eq.~(\ref{eqn:groundstates}). There are four flavors of
bosons corresponding to the distinct sublattices labelled with subscript $a$. 
One performs a Bogoliubov
 transformation taking boson operators $a_{a}^{\dagger}(\mathbf{k})$ and
$a_{a}(\mathbf{k})$ into spin wave creation and annihilation
operators $c_{A}^{\dagger}(\mathbf{k})$ and $c_{A}(\mathbf{k})$ so that the
Hamiltonian to harmonic order is brought to the form
\begin{multline}
 H[\bar{\phi}] = -NJS(S+1) + JS\sum_{\mathbf{k},A}
\epsilon_{A}(\mathbf{k}) \\ + JS\sum_{\mathbf{k},A} \epsilon_{A}(\mathbf{k})
{ c_A^{\dagger}(\mathbf{k}) c_A(\mathbf{k})  }
\end{multline}
where $\epsilon_{A}(\mathbf{k})$ are the spin wave energies. 
Further details of Holstein-Primakoff linear spin waves
on a pyrochlore lattice of spins can be found in Ref.~[\onlinecite{SpinWaves}]. 
As one would expect for an antiferromagnet, the
dispersion for the model Eq.~(\ref{eqn:model}) about the zero
modes is linear in $\vert\mathbf{k}\vert$. 
Just as in the classical case, the zero modes appear in pairs of planes
in the first Brillouin zone for the $\psi_{2}$ states.
The harmonic correction to the ground state energy is
\begin{equation}
N\Delta E_{0}[\bar{\phi}] \equiv 
\left ( \frac{NJS}{4} \times \frac{1}{4} \right )
{
\sum_{A}
\int _{\rm BZ} \frac{d^{3}\mathbf{k}}{(2\pi)^{3}} \epsilon_{A}(\mathbf{k}) 
}
 \end{equation}
which we have evaluated numerically. The results are shown in
Fig.~\ref{fig:ZP}. To harmonic order, one observes that, among the $\mathbf{q}=0$ ground states, 
the zero point energy is minimized at the
$\psi_{2}$ spin configurations ($\bar{\phi}_{a}=n\pi/3$ for all $a$), 
so a quantum order-by-disorder mechanism selects the same states as thermal fluctuations. 

\subsection{Monte Carlo results}
\label{sec:OBDb}

To confirm the thermal order-by-disorder mechanism argued for in
Section~\ref{sec:OBDa}, and to investigate further the concern,
expressed in Ref.~[\onlinecite{OBD2}], that
the $\psi_2$ long-range order might 
not survive in the thermodynamic limit, we performed Monte Carlo simulations of
the nearest neighbour exchange only model. Parallel tempering Monte Carlo simulation 
were carried out with $J=1$ (and $\mathcal{D}=0$) for four
different system sizes, $L=2,3,4$ and $5$, of $L^{3}$ cubic unit cells of $16$
spins. To equilibrate the system, $5\times 10^{6}$ Monte Carlo sweeps were performed,
followed by the same number of steps to collect data. 
All four system sizes were found to have equilibrated satisfactorily according to
the criteria discussed in Sec.~\ref{sec:methods}.
We note that we were unable to obtain well equilibrated
results for $L=6$ even using parallel tempering. 


\begin{figure}
\includegraphics[width=0.5\columnwidth]{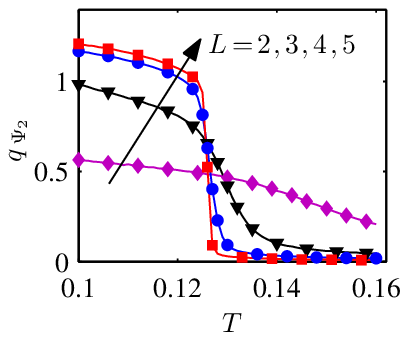}\includegraphics[width=0.5\columnwidth]{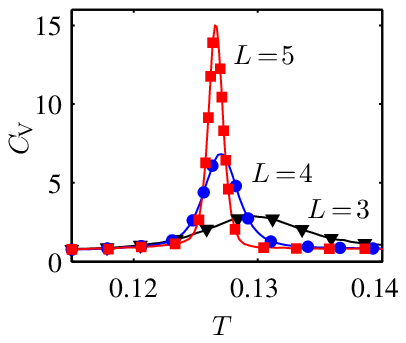}
\caption{(Color online) Order parameter $q_{\psi_{2}}$ and heat capacity $C_{\rm V}$ as a
  function of temperature for $J=1$ and $\mathcal{D}=0$. Left and right panels
  display results 
for $q_{\psi_{2}}$ ($L=2,3,4,5$) and for $C_{\rm V}$ ($L=3,4,5$), respectively.
}
\label{fig:D=0}
\end{figure}

\begin{figure}
\includegraphics[width=0.5\columnwidth]{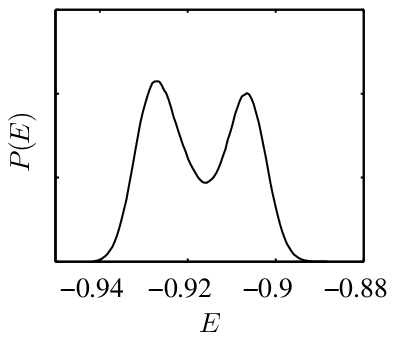}\includegraphics[width=0.5\columnwidth]{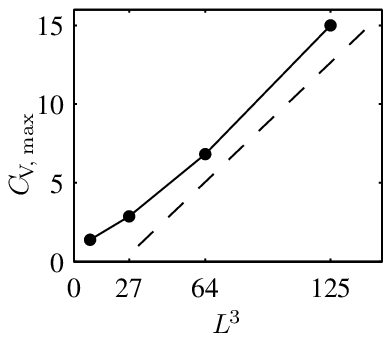}
\caption{(Color online) Two plots showing Monte Carlo data for $J=1$ and $\mathcal{D}=0$ illustrating
the first order nature of the transition. 
The left hand panel is a histogram of the
measured energies for $L=4$ at   $T/J\simeq 0.127$, close to the transition temperature. 
The double-peaked structure is evidence for a coexistence region and hence an underlying 
first order transition.
The right panel shows the peak height of the specific heat, $C_{\rm V,max}$, versus
the cube of the system size, $L^3$. 
The dash line shows a straight line for the hypothetical $C_{\rm V,max} \propto a+bL^3$ 
in the thermodynamic limit.
}
\label{fig2:D=0}
\end{figure}


\begin{figure}
\includegraphics[width=1\columnwidth]{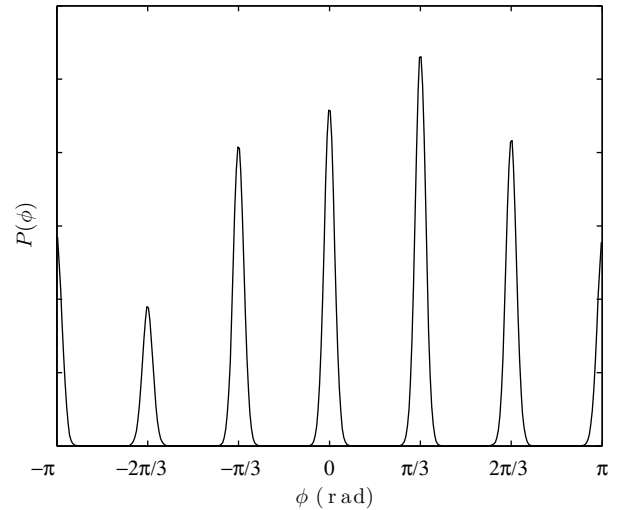}
\caption{
Monte Carlo data for $J=1$ and $\mathcal{D}=0$. 
Histogram of the local XY angles for all spins for $L=4$ at a temperature $T/J=0.1\lesssim T_{c}$. 
The histogram shows peaks at angles $n\pi/3$ illustrating that the $\psi_2$ states
are preferably sampled below $T_c/J \approx 0.127$.
}
\label{fig3:D=0}
\end{figure}


Figures \ref{fig:D=0}, \ref{fig2:D=0} and \ref{fig3:D=0} show data from the Monte Carlo simulations for
$\mathcal{D}/J=0$.
The left hand panel of Fig.~\ref{fig:D=0} shows the onset of the $\psi_{2}$ order parameter
while the right panel shows the temperature dependence of the specific heat, $C_V$, 
 near $T/J=0.127$, the estimated transition temperature, for various system sizes.
We have found that this estimated transition temperature 
$T_c/J \approx 0.127$ is consistent both with our
sublattice magnetization results (not shown) and those of Ref.~[\onlinecite{OBD2}]. 

 Both the rate of increase of the
heat capacity peak and the jump in
$q_{\psi_{2}}$ with increasing $L$ are consistent 
with a first order phase transition in the thermodynamic limit. 
The left hand panel of Fig.~\ref{fig2:D=0} is a histogram of the measured energies close to
the transition temperature for $L=4$. Its double-peaked structure is a clear
indication of co-existence and hence of the first order nature of the transition.
The right panel of Fig.~\ref{fig2:D=0} shows
how the peak height of the specific heat, $C_{\rm V,max}$, depends on the 
the cube of the system size, $L^3$. For a first order transition, one
expects $C_{\rm V,max} \propto (a+bL^3)$ in the limit of large $L$.~\cite{Binder}
The plot illustrates that for $L=4$ ($L^3=64$) and $L=5$ ($L^3=125$), 
$C_{\rm V,max}$ is approaching this expected behavior. This provides
 further evidence for a first order transition in this ${\cal D}=0$ $\langle 111\rangle$
pyrochlore XY antiferromagnet.
Finally,  Fig.~\ref{fig3:D=0} shows a histogram of the local XY
angle averaged over all spins on all sublattices at $T/J=0.1$ for system size
$L=4$. The figure shows six sharp peaks concentrated at the $\psi_{2}$ angles
$n\pi/3$. We find, in addition, that 
the spin angle on 
all sublattices are concentrated around
one of these angles at any given Monte Carlo time. This result therefore
demonstrates the selection of $\psi_{2}$ states from the continuous manifold
of classical ground states and 
also that all six magnetic domains are
sampled in the course of the simulation
 $-$ a possibility facilitated by the use of a parallel tempering
algorithm in our simulations compared to those of Refs.~[\onlinecite{OBD1,ETO,OBD2}]. 

\subsection{Further Cases of Order-by-Disorder in Pyrochlores}
\label{sec:otherwork}

We have described in detail the nature of the ground states in the 
$\langle 111\rangle$ XY pyrochlore antiferromagnet and how the classical degeneracy is 
resolved via an order-by-disorder mechanism. 
In contrast, the classical Heisenberg model on a
pyrochlore lattice, which has a much less constrained set of ground states, exhibits no phase
transition down to zero temperature.~\cite{MoessnerChalker}
To put our results in the broader context of order-by-disorder in pyrochlore systems, we
summarize in this short section
two previously studied cases where the degeneracy of the Heisenberg antiferromagnet can be lowered 
by including additional interactions, leading to
an entropic selection of a discrete state of long-range ordered states.

The first case is the pyrochlore Heisenberg model with both isotropic exchange
and Dzyaloshinskii-Moriya (DM) interactions~\cite{Elhajal,Chern2}
\[ 
H = J\sum_{\langle i,j\rangle} \mathbf{S}_{i}\cdot\mathbf{S}_{j} 
+ \sum_{\langle i,j\rangle} \mathbf{D}_{ij}\cdot(\mathbf{S}_{i}\times\mathbf{S}_{j})   .
  \]
Figure $2$ in Ref.~[\onlinecite{Elhajal}] gives the $\mathbf{D}_{ij}$ which are
completely determined by the lattice symmetry. We have taken a positive sign
in front of the DM term to denote the so-called indirect DM couplings of
Ref.~[\onlinecite{Elhajal}]. In this case, the single tetrahedron ground states
have four branches. One of these branches correponds exactly to the equal
$\bar \phi_{a}$ angle ground states of the $\langle 111 \rangle$ XY model [our
Branch 1 of Eq.~(\ref{eqn:groundstates})]. The other three branches are
coplanar spin configurations - the branches are distinguished by the three
mutually perpendicular normals to these planes in the $\langle 100\rangle $ crystallographic
directions. 
The authors of Ref.~[\onlinecite{Elhajal}] observed that Monte Carlo simulations at
low temperatures lead to the selection of a discrete state of states - breaking
down the 
continuous zero temperature classical degeneracy down to
a $Z_{6}$ symmetry. In the notation convention
of Ref.~[\onlinecite{ETO3}], these are the $\psi_{3}$ states. 
However, the nonzero
temperature ordering turns out to be more complicated. 
As observed very recently in Ref.~[\onlinecite{Chern2}], upon
lowering the temperature from the paramagnetic phase, there is a phase transition into
a $\psi_{2}$ long-range ordered phase followed, at a lower temperature, by the
$\psi_{3}$ ordering reported in Ref.~[\onlinecite{Elhajal}]. 
This finding was confirmed within a harmonic
Holstein-Primakoff computation of the free energy. \cite{Chern2}

The second example of order-by-disorder 
we discuss is the Heisenberg pyrochlore antiferromagnet 
with both nearest-neighbor
isotropic exchange and second neighbor
interactions. When the second neighbor exchange is ferromagnetic, the finite
temperature phase diagram explored by Monte Carlo \cite{Chern} exhibits an intermediate
phase that appears at intermediate temperatures between the collective
paramagnet and the low temperature incommensurate multiple-$\mathbf{q}$ ordered phase. 
This intermediate phase is partially ordered in the following sense. The magnetic
structure is layered $-$ each layer exhibiting collinear spins in one of three perpendicular
axes $\mathbf{n}$ which is common to all the layers. However, the
orientation of the spins along the $\mathbf{n}$ axis is apparently not
correlated between the layers. The partially ordered phase is selected
entropically as confirmed by a computation of the free energy.~\cite{Chern}
The latter analytical calculation  needs to, and does, include anharmonic terms
 to lowest order because, remarkably,
the entropically selected intermediate phase is not a local minimum 
of the harmonic approximation to the free energy.

\section{Materials Context}
\label{sec:eto}
 
\subsection{\eto}

One motivation for our interest in the local easy plane pyrochlore
antiferromagnet is the material \eto. The Curie-Weiss temperature
$\theta_{\rm CW}$ of this material is negative \cite{ETO7} varying between
$-13$ K and $-22$ K depending on the temperature range of measurement. \eto\
exhibits a transition at about $T_{c}=1.2$ K.~\cite{ETO} 
The $\psi_{2}$ magnetic structure of the ordered phase
proposed in Ref.~[\onlinecite{ETO}] has been confirmed by an analysis
of polarized neutron scattering data.\hspace{1pt}\cite{ETO3}

 The spin wave spectrum has been measured recently \cite{ETO5} revealing
 the presence of an almost gapless mode at $\mathbf{q}=0$ with a linear
 dispersion for small $\vert\mathbf{q}\vert$. 
This result is consistent with the $T^{3}$ magnetic specific
heat trend with temperature $T$. \cite{ETO,ETO8} The existence of a true Goldstone mode
  in this material is inconsistent with the selection of a discrete set of
  ordered states. The spin wave and specific heat 
  experimental results must therefore be providing an
  upper bound on the size of the 
spin wave excitation gap in this material
 Specifically, the inelastic neutron scattering
  spectrum in Ref.~[\onlinecite{ETO5}] can resolve the existence of a gap that is
  larger than about $1$ K and the heat capacity measurements of Ref.~[\onlinecite{ETO8}]
  show that the heat capacity varies as $T^3$ down to the base temperature of
  $450$ mK. One especially interesting feature of this material is that it apparently
exhibits a field-driven quantum phase transition.~\cite{ETO5} 
Application of a magnetic field in the $[110]$ direction
below $T_{c}$ causes a canting of the spins and leads to a zero mode in the
vicinity of $1.5$ T, which has been put forward as evidence for 
a finite field quantum phase transition in this material. \cite{ETO5,GTOFootnote}

A greater understanding of the quantum critical point is likely to rely 
on a deeper understanding of the microscopic mechanism responsible for
the zero field transition in this material.
One difficulty with the local XY pyrochlore antiferromagnet as an effective model
for \eto\ is that, unlike \eto, the model exhibits a strong first order
transition as we demonstrated in Section~\ref{sec:OBDb}. \cite{ETO} Another
crucial difficulty is that the dipolar coupling of Er$^{3+}$ ions is a
significant perturbation to the estimated exchange. \cite{ETO}
 The dipolar interaction renders the $\psi_{2}$ structure 
energetically unstable and favors a distinct ground state,
the $\psi_4$ (Palmer-Chalker) state,  as pointed
out originally in Ref.~[\onlinecite{ETO}]. 
In the next subsection, we
concentrate on the dipolar
interaction as a perturbation to isotropic exchange interactions. 
Specifically,
 we consider the possibility that thermal order-by-disorder might
persist in the presence of dipolar perturbations. 
After that, 
in Section \ref{sec:OBDb},
we summarize the theoretical constraints that have been placed on the problem of understanding
\eto\ in zero applied field.

\subsection{Effect of competing dipolar interaction}
\label{sec:competing}

In this section, we look at the effect of introducing pseudo-dipolar 
interactions on the order-by-disorder in the $\langle 111\rangle$ XY pyrochlore antiferromagnet.
 We find that order-by-disorder into $\psi_{2}$ states survives for sufficiently small 
pseudo-dipolar couplings and estimate the maximum value of the dipolar coupling strength ${\cal D}$
that permits $\psi_{2}$ ordering.

\begin{figure}
\includegraphics[width=0.5\columnwidth]{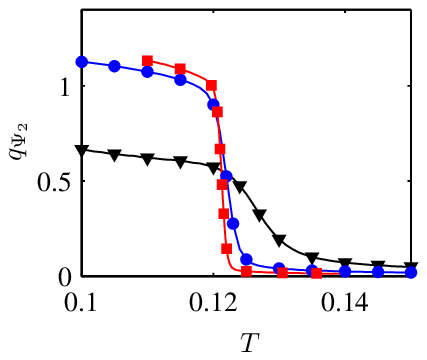}\includegraphics[width=0.5\columnwidth]{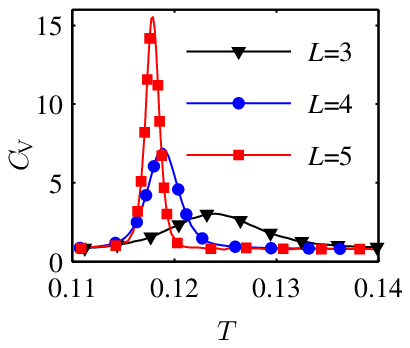}
\includegraphics[width=0.5\columnwidth]{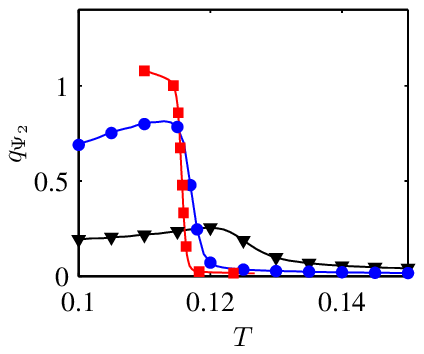}\includegraphics[width=0.5\columnwidth]{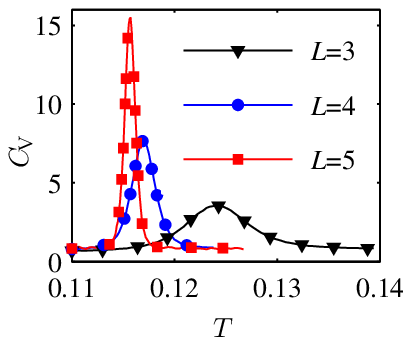}
\includegraphics[width=0.5\columnwidth]{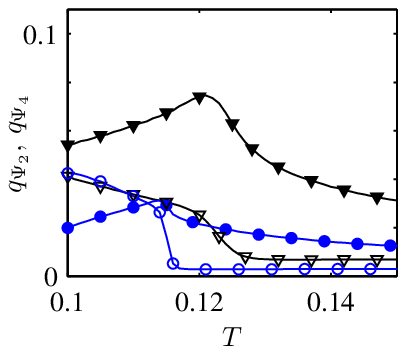}\includegraphics[width=0.5\columnwidth]{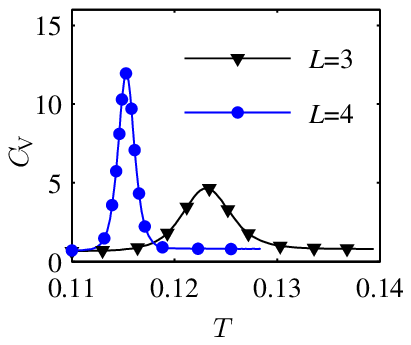}
\includegraphics[width=0.5\columnwidth]{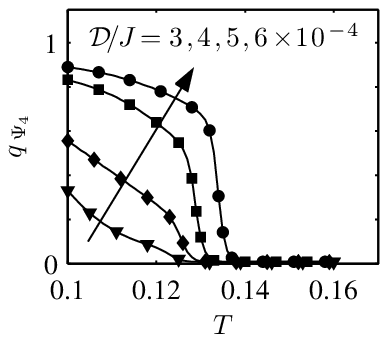}\includegraphics[width=0.5\columnwidth]{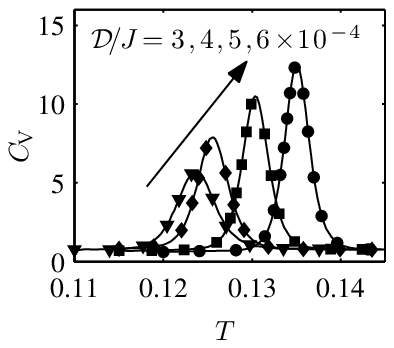}
\caption{(Color online) Figures showing the order parameters $q_{\psi_{2}}$ and
  $q_{\psi_{4}}$ and heat capacities for various values of
  $\mathcal{D}/J$. The top two rows, from top to bottom for
  $\mathcal{D}/J=0.5\times 10^{-4}$ and $\mathcal{D}/J= 10^{-4}$, show $q_{\psi_{2}}$ and heat capacity data for
  $L=3,4,5$. The third row down shows $C_{\rm V}$ and both $q_{\psi_{2}}$ (solid symbols) and
  $q_{\psi_{4}}$ (open symbols) for $\mathcal{D}/J= 2\times 10^{-4}$
  with $L=3$ (down triangles) and $L=4$ (circles) data.
 The bottom row of figures shows $q_{\psi_{4}}$ and
  heat capacity data for $L=3$ and $\mathcal{D}/J=3,4,5,6 \times 10^{-4}$, 
  with $\mathcal{D}$ increasing as indicated by the arrows.}
\label{fig:JDresults}
\end{figure}

The dipolar interaction in the following is taken to act solely between nearest neighbors in
order to reduce the difficulty in equilibrating the system and hence decrease
the computer simulation time.
We study this simplified model because (i) in combination with antiferromagnetic
exchange, dipolar interactions between nearest
neighbors select the same $\psi_{4}$ magnetic order as the long-range
dipolar interactions \cite{PalmerChalker} and 
because (ii) we have found, through preliminary simulations of Monte Carlo simulations of
a model with long-range dipolar interaction  (not reported),
that the general conclusions of this subsection below
do not depend sensitively on the range of these pseudo-dipolar interactions.

In order to examine the effect of dipolar perturbations on the thermal order-by-disorder mechanism,
 we carried out parallel tempering Monte Carlo simulations of the
model with Hamiltonian (\ref{eqn:model}) with nearest-neighbor dipolar
interactions. With the introduction of these pseudo-dipolar interactions, the simulations
were found not to equilibrate sufficiently for system sizes $L>5$ for
$\mathcal{D}/J\lesssim 10^{-4}$ and poor equilibration was found for $L>4$ for
$\mathcal{D}/J\gtrsim 10^{-4}$. The results presented in this section were obtained
 from simulations with $10^{8}$ Monte Carlo sweeps of which the last $1/10$th was
used to compute the thermal averages after equilibration. Increasing the
length of the simulations to $10^{9}$ Monte Carlo sweeps was found neither to
improve the results given below nor to achieve equilibration for larger system
sizes than those presented below. The simulations were performed on 
SHARCNET using the Saw cluster with Intel Xeon 2.83GHz processors, with the $L=5$ simulations
requiring $10^{8}$ Monte Carlo sweeps consuming $\sim 17$ CPU hours. Each
simulation was run on $64$ processors $-$ one processor for each temperature.

Figure \ref{fig:JDresults} shows the heat capacities and order parameters
$q_{\psi_{2}}$ and $q_{\psi_{4}}$ for
different values of $\mathcal{D}/J$
(see Section \ref{sec:methods} for a definition of $q_{\psi_{2}}$ and $q_{\psi_{4}}$).
The two top panels (left and right)  show the
results for $\mathcal{D}/J=0.5\times 10^{-4}$ for $L=3,4$ and $5$. There is a clear
onset of the $\psi_{2}$ order parameter upon lowering the temperature and
this feature becomes {\it sharper} when the system size increases. Similar behavior is shown in
the second row down in Fig.~\ref{fig:JDresults} for a larger value,
$\mathcal{D}/J=10^{-4}$. The $\psi_{4}$ order
parameter was also measured for these two values of $\mathcal{D}/J$ and found to be close to zero,
within the present finite size effects,
and displaying no perceptible features around the onset temperature of $q_{\psi_{2}}$. This
is, therefore, compelling evidence for 
the thermal order-by-disorder
 of an ordered $\psi_2$ state at  nonzero temperature
 persisting in the presence of
pseudo-dipolar interactions that break the ground state degeneracy in such a way that the thermally
selected states are not the true ground states. The
figures in the third row down (for $\mathcal{D}/J=2\times 10^{-4}$) 
reveal strong competition between
$\psi_{2}$ and $\psi_{4}$ ordering  as
indicated by the fact that both order parameters are finite but small and
suppressed upon increasing the system size. We were unable to obtain a clear
signature of any long-range order in this region of pseudo-dipolar coupling in
spite of the presence of a heat capacity peak that sharpens with increasing
system size (right-hand panel, third row down). The
bottom left-hand panel shows the $\psi_{4}$ order parameter for increasing
$\mathcal{D}/J$ showing the growing robustness of the transition to $\psi_{4}$
order as the pseudo-dipolar coupling is increased. 
For these values of $\mathcal{D}/J$, despite long runs of $10^9$ Monte
Carlo steps, the largest system size we were able to equilibrate was $L=3$.
 
The dipolar coupling of nearest neighbor Er$^{3+}$ 
moments in \eto\ is $\mathcal{D}/(2\sqrt{2})^{3}=D_{\rm
  nn}=\mu_{0}(g_{J}\mu_{B})^{2}/4\pi R_{\rm nn}^{3}=0.02$ K.
 Whether other (anisotropic) exchange interactions \cite{ETO6} 
acting between the Er$^{3+}$ ions are small compared to the isotropic exchange is
not known at this time (see Ref.~[\onlinecite{YTO}] for a discussion of this problem for
the Yb$_2$Ti$_2$O$_7$ XY pyrochlore).

Nevertheless, a fit to the local susceptibility of \eto\ in a
$[110]$ magnetic field of $1$ T (see Ref.~[\onlinecite{Cao}]) has yielded an
order of magnitude estimate for the exchange coupling of about $10^{-1}$ K giving $(D_{\rm
  nn}/J)_{\rm exp} \sim 0.2$. 
On the other hand, Monte Carlo simulations gives an approximate boundary between the
 $\psi_{2}$ and $\psi_{4}$ long-range ordered phases of about $\mathcal{D}/J\sim 2\times10^{-4}$ (see
Fig.~\ref{fig:JDresults}) or, in other words, a condition for the
appearance of the $\psi_{2}$ phase of $D_{\rm nn}/J \lesssim 0.005$. 
Therefore, one is faced with the following conundrum.
For the experimental strength of the dipolar coupling and the estimated exchange coupling,
 the ordering transition in Er$_{2}$Ti$_{2}$O$_{7}$ should, according to our Monte
Carlo simulations, take place
into  $\psi_{4}$ (Palmer-Chalker) states $-$  but an ordering into 
$\psi_{2}$ states is observed experimentally.
This highlights a problem with the thermal order-by-disorder scenario for Er$_2$Ti$_2$O$_7$.

\subsection{Towards a model for \eto}

It is perhaps worth summarizing the physics problem associated with the zero
field ordered state of \eto. 
The  long-range ordered magnetic structure 
that is experimentally observed in \eto\ arises via thermal
 order-by-disorder in the local $\langle 111\rangle$ XY antiferromagnet,
as discussed in Section \ref{sec:OBD}.
A large $S$ expansion indicates, furthermore, that there is
quantum order-by-disorder with the result that the
$\psi_{2}$ states are ground states of this model. However, as we have shown
in Section \ref{sec:competing},
the addition of a dipolar interaction (which is known to be present and
relatively large in Er$_2$Ti$_2$O$_7$) is
inconsistent with the selection of $\psi_{2}$ ordered states in the classical
model. In order to resolve this paradox we might be led to consider the
following possibilities: (i) quantum order-by-disorder overcomes the dipolar interaction as
conjectured by Ref.~[\onlinecite{ETO}], (ii) the nearest-neighbor bilinear exchange interaction is
 anisotropic \cite{ETO6,YTO,Cao} and uniquely energetically
selects a $\psi_2$ ordered state, (iii) there  exist multipolar interactions
between the Er$^{3+}$ ions accounted for neither in the present work nor
in previous ones, \cite{Multipolar} (iv) the crystal field is responsible for the
effective anisotropy stabilizing the $\psi_{2}$ long-range ordered
state \cite{ETO6} and (v) that further neighbor interactions might be
important in this problem.

It has been
shown that allowing for anisotropic
exchange interactions does not lessen the conceptual difficulties in
understanding the $\psi_{2}$ ordering in \eto. \cite{ETO6} 
Specifically, as discussed in Ref.~[\onlinecite{ETO6}],
 no bilinear nearest-neighbor interactions in the idealized
  $\langle 111\rangle$ XY model on a 
 pyrochlore can bring about unique classical zero temperature $\psi_{2}$ ground states.
The $\psi_{2}$ structure can be obtained, however, via local
mean field theory by considering the full crystal field spectrum in addition to
the anisotropic exchange. In this case,
the effective single ion sixfold anisotropy arises from interaction-induced admixing of excited
crystal field wavefunctions into the single ion ground state
doublet. \cite{ETO6} 
Unfortunately, even in this case, the strong long
range dipolar interaction 
between Er$^{3+}$ moments in \eto\
is expected to drive the system into a
different magnetically ordered (Palmer-Chalker) state, $\psi_{4}$.~\cite{PalmerChalker} 
This is because the energetic selection produced
by the dipolar interaction inherent to the Er$^{3+}$ magnetic
 moments outweighs the six-fold single ion
anisotropy effect above 
(that perturbs the XY-like single ion ground state doublet) by at least two
orders of magnitude. The weakness of the $\psi_{2}$ selection 
proceeding via the involvement of the excited crystal field states
 is due to the large size of the crystal field gap ($\sim 10^2$ K \cite{ETO})
compared to the perturbing interactions.~\cite{ETO6} 
In short, the combination of anisotropic exchange and
crystal field effects do not appear to be able to win against the strong
dipolar interactions and 
{\it energetically}
stabilize the experimentally 
observed $\psi_{2}$ \cite{ETO,ETO3} state rather than the
$\psi_{4}$ (Palmer-Chalker) ~\cite{PalmerChalker} state. 
This would seem to rule out possibilities (ii) and (iv).

In addition, the existence of a ground state crystal field
doublet separated from the first excited states by a gap much larger than the
scale of the interactions implies that multipolar interactions
between pairs of angular momenta in the microscopic model map to anisotropic
 bilinear exchange couplings \cite{ETO6}
in the low energy effective spin-1/2 theory,~\cite{Yb2Ti2O7-example} thus 
seemingly ruling out
multipolar interactions as the microscopic mechanism behind the zero field
long-range order in \eto. 
\footnote{In more detail, suppose that the
  microscopic interactions between magnetic ions are two-body interactions
  that respect the lattice symmetries. Suppose, furthermore, that the crystal
  field spectrum has a ground state doublet and a gap to excited single ion states,
  $\Delta$, that is much larger than the energy scale associated with the
  interactions. Finally, we suppose that we are interested in temperatures
  $T\ll \Delta$. Then the excited crystal field can be neglected and the
  effective theory at low energies is the projection of the microscopic
  Hamiltonian onto the product of crystal field doublets on each magnetic ion.
Since the space of states at each magnetic site is two dimensional, all
site operators are linear combinations of Pauli spin operators regardless of
the nature of the bare microscopic interactions. Therefore multipolar
interactions project onto bilinear exchange couplings.} 
Further neighbor interactions (which can also be anisotropic) have not yet been investigated.


We might consider instead a compromise between anisotropic exchange and
thermal order-by-disorder $-$ that important nearest-neighbor exchange anisotropy
might more than cancel off the nearest-neighbor 
part of the interaction coming from the long-range dipolar interaction.
The local XY antiferromagnet in the presence of pseudo-dipolar interactions
with a coupling of the opposite sign 
[i.e. ${\cal D}<0$ in  Eq.~(\ref{eqn:model})]
to the physical long-range dipolar
interaction preserves a continuous degeneracy that includes the discrete
$\psi_{2}$ states.~\cite{ETO6} 
However, the degeneracy is reduced from the
case with only exchange interactions.
We might naively expect that the
$\psi_{2}$ states will   generically cease to be
 the configurations 
with the highest density of soft modes when such anisotropic
interactions are introduced and hence that thermal order-by-disorder may 
 not occur, at least not along the lines of the
specific mechanism discussed in Section \ref{sec:OBDa}.
However, this naive reasoning may not be correct in general,
as we now briefly mention.

An exception that was noted
recently \cite{Chern2} is the pyrochlore Heisenberg antiferromagnet with
Dzyaloshinskii-Moriya interactions.
As mentioned in Section~\ref{sec:otherwork}, Monte Carlo simulations and
calculations of the free energy for that model
reveal that $\psi_{2}$ states are entropically
selected at intermediate temperatures between the high temperature paramagnet
and a low temperature coplanar state. The effect of dipolar interactions 
was not investigated in Ref.~[\onlinecite{Chern2}]. However, one may speculate 
that in this, or some other model with anisotropic exchange, 
in the presence of dipolar interactions, $\psi_2$ states are selected
 as an intermediate temperature phase which remains as a metastable
 phase upon cooling. In other words, only one transition might be observed 
although two would be observed if the system were able to equilibrate.

The remaining possibility -- that the $\psi_{2}$
states are the ground states of the quantum model 
despite strong dipolar interactions classically stabilizing a
Palmer-Chalker $\psi_{4}$ state
 -- is one that has not yet been investigated.
 However, in the presence of the
resulting strong and necessarily nonlinear (e.g. beyond $1/S$)
quantum fluctuations,
required 
to overcome the dipole-induced $\psi_4$ ground state,
one might expect a substantial reduction of the ordered moment
compared to the moment of the noninteracting doublet.
This expectation contrasts with the
experimental situation where the
ordered moment is about $3 \mu_{B}$ \cite{ETO} compared to a
noninteracting moment of $3.8 \mu_{B}$ -- a mere $20 \%$ reduction. Still, such
an observation perhaps does not rule out
quantum order-by-disorder which may 
therefore remains a possible solution to the problem of zero field ordering in \eto\
as originally conjectured in Ref.~[\onlinecite{ETO}]. 
We hope that a future study will
investigate the problem of the quantum ground state
of the local $\langle 111\rangle$ XY antiferromagnet in the
presence of dipolar interactions perhaps by carrying out an anharmonic $1/S$
calculation of the ground state energy. Meanwhile, the microscopic mechanism
that gives rise to a $\psi_{2}$ ordered phase at $1.2$ K and with seemingly
small quantum zero point fluctuations in \eto\ remains an open and interesting
problem in the field of high frustrated magnetism.

In summary, no mechanism has yet been identified that can simultaneously explain
how the (dipole-driven) energetic selection of the $\psi_{4}$ state is avoided in \eto while allowing
for selection of the $\psi_{2}$ state below a critical temperature $T_c \sim 1.2$ K.
 This may indicate that quantum
fluctuations do play a crucial role in the $\psi_{2}$ ordered state observed
in \eto\ and that a renewed investigation of their effects is warranted.

\section{Summary}

We have presented in Sections~\ref{sec:model} and \ref{sec:OBD}, in some detail, the
classical ground states, thermal and quantum behavior of the pyrochlore $\langle
111\rangle$ XY antiferromagnet with exchange only. We have shown that the classical ground states
on a single tetrahedron have four branches of ground states each with one
continuous degenerate set of states involving the smooth rotation of all four
spins simultaneously, confirming the previous result.~\cite{OBD2}
From a calculation of the ground states on a single tetrahedron, we have
inferred the ground states on the pyrochlore lattice which include line
defects implying that the number of ground states scales as $L^{2}$,
 where $L$ is the edge length of the crystal.

Monte Carlo simulations of this model confirm that there is thermal selection
of a discrete set of spin states, denoted $\psi_{2}$ states, \cite{ETO3}
 with ordering wavevector $\mathbf{q}=0$ from
the manifold of classical ground states. We have shown, furthermore, that this
selection occurs to harmonic order in small angular fluctuations about the
classical ground states. In this model, the linear spin wave spectrum shows
strong similarities with the spectrum of eigenvalues of the Hessian.
Specifically,
 the spin wave zero modes and the Hessian zero modes appear
 within the same planes in reciprocal space for the $\psi_2$ states. 
It follows that the quantum
zero point energy is minimized at the same $\psi_{2}$ spin configurations that are selected
through a thermal order-by-disorder mechanism,
 hence confirming the conjecture of Refs.~[\onlinecite{ETO,OBD2}].


We have considered the effect of including (nearest-neighbor) 
pseudo-dipolar interactions together with
the antiferromagnetic exchange in the classical model at finite
temperature. The (energetically selected) ground states of this model are the
$\psi_{4}$ states \cite{ETO3} (also referred to as the Palmer-Chalker state
\cite{PalmerChalker}), given in Section~\ref{sec:model}, 
so the introduction of dipolar interactions 
produces a competition between energetic selection and
thermal selection. We have found evidence 
for the persistence of an order-by-disorder transition 
to a  $\psi_{2}$ state even when $\mathcal{D}\neq 0$. 
This finding
implies that, in principle, a second transition 
should occur at lower temperatures into
the  $\psi_{4}$ 
(Palmer-Chalker) long-range ordered state since it is the
classical ground state.
However, using Monte Carlo simulations, we have found no evidence for such a transition, at
the very least, because the difficulties of equilibration within the ordered
$\psi_{2}$ phase prevent the exploration of the space of configurations
computationally.
A similar situation arises in a model that tunes between the
Heisenberg pyrochlore antiferromagnet and the fcc Heisenberg antiferromagnet
in which thermal order-by-disorder, studied using Monte Carlo simulations,
prevails over the energetically driven ordering identified
within mean field theory.~\cite{Pinettes} 
No transition was reported in Ref.~[\onlinecite{Pinettes}] 
from the entropically selected ordered phase
to the energetically favoured magnetic ordered phase. In contrast, the high
and low temperature phase boundaries of an entropically stabilized
intermediate (finite temperature) state have
been identified in the $J_{1}/J_{2}$ pyrochlore Heisenberg antiferromagnet. \cite{Chern}

To conclude, we have shown that $\psi_2$ long-range order is present at low temperatures 
in the $\langle 111\rangle$ XY pyrochlore
antiferromagnet induced both by thermal and quantum fluctuations.
How such $\psi_2$ states, either as a zero temperature ground state or as an ordered state at
nonzero temperature, arise in the 
presence of (long-range) dipolar interactions remains an
intriguing question that will require further theoretical investigation.

\begin{acknowledgments}
We thank Bruce Gaulin, Peter Holdsworth, Kate Ross, Jacob Ruff
and Jordan Thompson for useful discussions. 
This research was funded by the NSERC of Canada and the Canada
Research Chair program (M. G., Tier I). We acknowledge the use of
the computing facilities of
the Shared Hierarchical Academic Research Computing Network (SHARCNET:www.sharcnet.ca).
\end{acknowledgments}

\appendix

\section{Calculation of the Ground States of the XY Antiferromagnet}

The zero moment conditions on a single tetrahedron are
\begin{align}
&\cos\left(\phi_{1}\right) + \cos\left(\phi_{2}\right) = \cos\left(\phi_{3}\right) + \cos\left(\phi_{4}\right) \\
&\cos\left(\phi_{1}'\right) + \cos\left(\phi_{3}'\right) = \cos\left(\phi_{2}'\right) + \cos\left(\phi_{4}'\right) \\
&\cos\left(\phi_{1}''\right) + \cos\left(\phi_{4}''\right) = \cos\left(\phi_{2}''\right) + \cos\left(\phi_{3}''\right),
\label{eqn:zeromoment_app}
\end{align}
\noindent where $\phi_{a}'\equiv \phi_{a}+\frac{2\pi}{3}$ and $\phi_{a}''\equiv
\phi_{a}+\frac{4\pi}{3}$. The angles $\phi_{a}$ are angles in the local
coordinate system on sublattice $a$ ($a=1,2,3,4$).
As discussed in Section ~\ref{sec:model}, and as previously reported in Ref.~[\onlinecite{OBD3}],  the ground
states of the $\langle 111 \rangle$ XY pyrochlore antiferromagnet are characterized by the following
four set of solutions (``branches''):
\begin{align}\bar
{\rm Branch \hspace{1pt}1:}\hspace{6pt} \bar\phi & \equiv \bar\phi_{1} = \bar\phi_{2} = \bar\phi_{3} = \phi_{4} \notag \\
{\rm Branch \hspace{1pt}2:}\hspace{6pt} \bar\phi & \equiv \bar\phi_{1} = \bar\phi_{2} = -\bar\phi_{3} = -\bar\phi_{4} \notag \\
{\rm Branch \hspace{1pt}3:}\hspace{6pt} \bar\phi & \equiv \bar\phi_{1} = \bar\phi_{3}\hspace{1pt}, \hspace{4pt} \frac{2\pi}{3} - \bar\phi = \bar\phi_{2} = \bar\phi_{4} \notag \\
{\rm Branch \hspace{1pt}4:}\hspace{6pt} \bar\phi & \equiv \bar\phi_{1} = \bar\phi_{4}\hspace{1pt}, \hspace{4pt} \frac{4\pi}{3} - \bar\phi = \bar\phi_{2} = \bar\phi_{3}. 
\label{eqn:groundstates_app}
\end{align}

In this section, we present a derivation of this result that differs from the one in Ref.~[\onlinecite{OBD3}].
Defining $\sigma_{ab}\equiv (\phi_a + \phi_b)/2$ and $\delta_{ab} \equiv (\phi_a - \phi_b)/2$, we
first proceed to rewrite the zero moment conditions as 
\begin{align}
         \cos(\sigma_{12})\cos(\delta_{12}) -         \cos(\sigma_{34})\cos(\delta_{34}) = 0 \label{A} \\
     -   \sin(\sigma_{12})\sin(\delta_{12}) + \sqrt{3}\cos(\sigma_{34})\cos(\delta_{34}) = 0 \label{B} \\
\sqrt{3} \cos(\sigma_{12})\sin(\delta_{12}) -         \sin(\sigma_{34})\sin(\delta_{34}) = 0 \label{C} 
\end{align}
by using half-angle formulae and then combining the second and third zero
moment conditions in Eq.~(\ref{eqn:zeromoment_app}) to finally obtain Eqs.~(\ref{B}) and (\ref{C}).
Our strategy is to eliminate the sums of pairs, 
$\sigma_{\mu\nu}$, keeping only the differences of pairs, $\delta_{\mu\nu}$.
Thus, from Eq.~(\ref{A}), we get
 \[ 
\sin^{2}(\sigma_{12}) = 1 
- \left( \frac{\cos(\sigma_{34}) \cos(\delta_{34})}{\cos(\delta_{12})} \right)^{2}
 \]


\noindent Substitute into Eq.~(\ref{B}) to get
\begin{multline} 
\left(1-\sin^{2}(\delta_{12}) - \cos^{2}(\delta_{34})\cos^{2}(\sigma_{34}) \right)\sin^{2}(\delta_{12}) \\ =
3\cos^{2}(\sigma_{34})\sin^{2}(\delta_{34})\left(1-\sin^{2}(\delta_{12}) \right)
 \label{eqn:mult} 
\end{multline}


\noindent  Then,  squaring Eq.~(\ref{B}) and Eq.~(\ref{C}), and adding the result, we obtain
\begin{equation} 		
\sin^{2}(\delta_{12}) = \left(  \frac{1}{3} \sin^{2}(\sigma_{34}) + 3\cos^{2}(\sigma_{34}) \right) 
\sin^{2}(\delta_{34})
\label{Iden}  
\end{equation}
which we can otherwise write as
\begin{equation}
\cos^{2}(\sigma_{34}) = \frac{3}{8} \left(
      \frac{\sin^{2}(\delta_{12})}{\sin^2(\delta_{34})} -\frac{1}{3} \right)
\label{eqn:elim}
\end{equation}

So now we can proceed with what we set out to do: substitute Eq.~(\ref{eqn:elim}) into
Eq.~(\ref{eqn:mult}) leaving us, after re-arranging and cancelling off 
a $\cos^{2}(\delta_{12})$ term, with
\begin{multline}
\sin^{2}(\delta_{12}) \sin^{2}(\delta_{34}) = \\
\frac{1}{8}
\left[ 3\sin^{2}(\delta_{12}) - \sin^{2}(\delta_{34}) \right ]
\left[ 3\sin^{2}(\delta_{34}) + \sin^{2}(\delta_{12}) \right ]
\end{multline}
It follows from this last equation that
\[ \frac{3}{8}\left (  \sin^2(\delta_{34}) - \sin^2(\delta_{12}) \right) =0 . \]
Hence, the most general form for the set of $\phi_{a}$ angles is
\[ (\bar\phi_{1},\bar\phi_{2},\bar\phi_{3},\bar\phi_{4}) = (\bar\phi+\theta,\bar\phi,\psi\pm \theta,\psi) .    \]
Substituting this into our original zero moment formula, Eq.~(\ref{A}), which we write here again:
 \[ \cos\left(\bar\phi_{1}\right) + \cos\left(\bar\phi_{2}\right) 
= \cos\left(\bar\phi_{3}\right) + \cos\left(\bar\phi_{4}\right) , \]
shows that one must have either
\[ (\bar\phi_{1},\bar\phi_{2},\bar\phi_{3},\bar\phi_{4}) = (\phi+\theta,\phi, \phi+ \theta,\phi)    \]
or 
\[ (\bar\phi_{1},\bar\phi_{2},\bar\phi_{3},\bar\phi_{4}) = (\phi+\theta,\phi,\phi,\phi+\theta)     \]
if $\theta$ is nonvanishing or
\[ (\bar\phi_{1},\bar\phi_{2},\bar\phi_{3},\bar\phi_{4}) = (\phi,\phi,\pm\phi,\pm\phi)  \]
if $\theta=0$.
Thus, the angles must occur in pairs. 
One can now return to original zero moment conditions, Eq.~(\ref{eqn:zeromoment_app}), and, 
with the knowledge that the angles must occur in pairs, obtain the four branches of ground states in 
Eq.~(\ref{eqn:groundstates_app}). For example, suppose that the pairs of angles occur in the configuration
\[  \left(\psi,\phi,\phi, \psi \right)    \]
then Eq.~(\ref{eqn:zeromoment_app}) gives  the branch
\[  \left(\psi,\frac{4\pi}{3}-\psi,\frac{4\pi}{3}-\psi, \psi \right).      \]

\section{Brillouin Zone Integration}
\label{app:factors_of_four}

In the calculation of the free energy in Eq.~(\ref{eqn:FE}), in particular the last term of
that equation,  we take the Brillouin zone sum
over to an integral as applicable to the case of an infinite lattice. 
In general, we expect
\[ \sum_{\mathbf{k}} \rightarrow \frac{N_{\rm P}}{\Omega_{\rm BZ}} \int_{\rm BZ},     \]
where the integral is taken over the Brillouin zone of volume $\Omega_{\rm
  BZ}$ and $N_{\rm P}$ is the number of primitive cells. 
We thus have
\[  \sum_{\mathbf{k}} \rightarrow \frac{Na^{3}}{4(4(2\pi)^{3})}\int_{\rm BZ} d^{3}k.       \]
The edge length of the cubic unit cell, $a$, has been set equal to one. 
The two factors of one quarter come about because
 (i) $4N_{\rm P} = N$
where $N$ is the number of spins and (ii) the Brillouin zone volume is $4(2\pi/ a)^{3}$.

\section{Equipartition argument}
\label{sec:AvE}

In this section, we revisit the Monte Carlo simulations presented in Ref.~[\onlinecite{OBD2}]. 
In particular, we consider Fig.~\ref{fig:JDresults} in that work showing the average energy 
in the ordered phase at low temperatures as a function of the system size. 
The authors of Ref.~[\onlinecite{OBD2}] found that
\[    \frac{E}{Nk_{\rm B}T} = \alpha - \beta\frac{1}{L}    \]
 We find that the existence of chains of zero modes in the $\langle 111\rangle$ XY 
pyrochlore antiferromagnet is sufficient to constrain the coefficients $\alpha$ and $\beta$.

Let us consider a cubic cell with edge length $L$ with each cubic unit cell of unit edge length.
 The number of spins is $N=16L^{3}$.
Consider a square face of a single cubic cell. 
There are two chains beginning on the face that alternate between sublattices $a$ and $b$. 
The number of such chains passing through the sample is $N_{\rm chains}=2L^{2}$. 
There is one degree of freedom per spin so the average energy would be $(1/2)NkT$ 
were it not for the fact that the spectrum of modes about the $\psi_{2}$ states has planes 
of zero modes. We shall assume that the zero modes (in the harmonic spectrum) 
are actually resolved by a quartic contribution to the energy when looking at 
the higher order corrections to the Hamiltonian. The zero modes correspond 
to rotations along two classes of $ab$ chains so the number of such modes is $2N_{\rm chains}$. 
Thus, the average energy is
\[ \frac{E}{k_{\rm B}T} = \frac{1}{2}\left(N - 
2N_{\rm chains} \right) + \frac{1}{4}\left(2N_{\rm chains} \right) 
= \frac{1}{2}N - \frac{1}{2}N_{\rm chains}. \]
Then, because $N_{\rm chains}=2L^{2}=2(N/16)^{2/3}$, we obtain
\[  \frac{E}{Nk_{\rm B}T} = \frac{1}{2} - \frac{1}{16^{2/3}}N^{-1/3} 
 =  \frac{1}{2} - \frac{1}{16}\frac{1}{L} . \]
The coefficient in front of $1/L$ from Monte Carlo simulation \cite{OBD2}
 is approximately $0.0636$ which is in good
agreement with the calculated $1/16=0.06250$ value above.


\begin{thebibliography}{99}
\bibitem{Footnote} Refs.~[\onlinecite{OBDmodel1,OBDmodel2,OBDmodel3}] 
discuss a small assortment of models that exhibit order-by-disorder of some variety.
\bibitem{integrable}
The free energy is a smooth function of the parameter that tunes between
classical ground states despite the presence of zeros in the eigenvalues of
the Hessian. This is because the zeros occur on planes in the Brillouin
zone. Consider a plane of zeros in the yz plane with linear dispersion (for an
antiferromagnet) giving a $k_{x}$ dispersion for small $k_{x}$ away from the
plane of zeros. The free energy is, roughly, the integral over the Brillouin
zone of $\log k_{x}$ or 
\[ \int_{0}^{c} dk_x \log \vert k_x \vert   \]
where the upper (positive) limit is arbitrary. This integral is finite.
\bibitem{GTOFootnote} We note, however,  that similarly to the
  behavior of Gd$_{2}$Ti$_{2}$O$_{7}$ in a field \cite{GTO}, 
perhaps also in Gd$_2$Sn$_2$O$_7$ \cite{GSO-field},
it would seem plausible
  that \eto\ many exhibit several field driven transitions from its zero field
state to its highly polarized state at $\gtrsim 5-10$ T (see Ref.~[\onlinecite{ETO5}]).
For example, a calculation that considers the effect of a 
$[110]$ field in a microscopic model of \eto\ does find evidence for
multiple field-driven transitions (see Ref.~[\onlinecite{ETO6}]).
\end{thebibliography}
\end{document}